\documentclass[5p,sort&compress]{elsarticle}

\usepackage{hyperref}
\usepackage{amsmath}
\usepackage{amssymb}
\usepackage{slashed}
\usepackage{color}
\usepackage{array}

\journal{Physics Letters B}

\makeatletter
  \def\ps@pprintTitle{%
  \let\@oddhead\@empty
  \def\@oddhead{\hspace*{\fill} WSU-HEP-1801}%
  \def\@oddfoot{}%
  \let\@evenfoot\@oddfoot}
\makeatother

\begin{document}

  \begin{frontmatter}

    \title{Studies of Lepton Flavor Violation at the LHC}

    \author[inst1,inst2]{Bhubanjyoti Bhattacharya} \ead{bbhattach@ltu.edu}
    \author[inst2,inst3]{Robert Morgan} \ead{robert.morgan@wisc.edu}
    \author[inst2]{James Osborne\corref{cor1}} \ead{jaosborne@wayne.edu}
    \author[inst2]{Alexey A. Petrov} \ead{apetrov@wayne.edu}

    \address[inst1]{Department of Natural Sciences, Lawrence Technological University, Southfield, MI 48075, USA}
    \address[inst2]{Department of Physics and Astronomy, Wayne State University, Detroit, MI 48201, USA}
    \address[inst3]{Department of Physics, University of Wisconsin--Madison, Madison, WI 53706, USA}

    \cortext[cor1]{Corresponding author}

    \begin{abstract}
      We examine the charged lepton flavor violating process $gg \rightarrow \mu^\pm \tau^\mp$ at the $\sqrt{s} = 13$~TeV LHC. Operators generating this process can be induced by new physics (NP) at dimension 8. Despite the power suppression associated with dimension 8 operators, we show that the LHC's large gluon luminosity makes it possible to probe this channel. For an integrated luminosity of 100~fb$^{-1}$ at the LHC, we predict a constraint on the NP scale $\Lambda \gtrsim 3$~TeV. In addition, we point out that such operators can be induced through top quark loops in models that generate dimension 6 operators of the form $\overline{t} t \, \mu \tau$. We find that the NP scale of these dimension 6 operators can be constrained to be $\Lambda \gtrsim 3.4$--$4.1$~TeV with 100~fb$^{-1}$ of data.
    \end{abstract}

  \end{frontmatter}

  \section{Introduction}
  \label{sec:introduction}

  Lepton flavor violation (LFV) is an important vehicle for low energy studies of physics beyond the Standard Model (BSM). Within the Standard Model (SM) with massless neutrinos, individual lepton number is conserved. Even with the addition of non-zero neutrino masses, processes that violate charged lepton number are suppressed by powers of $m_\nu^2 / m_W^2$~\cite{Raidal:2008jk}. Thus, experiments should be extremely sensitive to BSM physics that facilitate charged lepton flavor-violating (CLFV) processes.

  SM phenomena are also expected to closely obey lepton-flavor universality (LFU). However, recent observations from LHCb~\cite{Aaij:2014ora,Aaij:2015yra,Aaij:2017vbb}, BaBar~\cite{Lees:2013uzd}, and Belle~\cite{Huschle:2015rga,Wehle:2016yoi} show hints of LFU violation in semi-leptonic decays of $B$ mesons at the level of a few standard deviations. In response to these findings there have been many proposals introducing new physics, for example studies of $b \rightarrow s \mu \mu$~\cite{Descotes-Genon:2013wba,Altmannshofer:2013foa,Gauld:2013qja,Datta:2013kja,Buras:2013dea} and lepton-flavor non-universal interactions~\cite{Altmannshofer:2014cfa,Sakaki:2013bfa} (for a recent review, see Ref.~\cite{Buttazzo:2017ixm} and references therein). Although not required~\cite{Celis:2015ara,Alonso:2015sja}, new interactions that violate LFU may also induce LFV~\cite{Glashow:2014iga}. With the prospects of studying LFV in $B$-meson decays at LHCb and the upcoming Belle II experiment, there has been renewed theoretical attention to this type of new physics~\cite{Bhattacharya:2014wla,Bhattacharya:2016mcc,Alok:2017jgr,Alok:2017sui,Altmannshofer:2017yso,Crivellin:2017zlb,Iguro:2017ysu,Iguro:2018qzf}.

In addition to studies of LFV in $B$ decays, some authors have proposed refined methods for direct searches at the Large Hadron Collider (LHC) to look for new TeV-scale particles that can mediate LFV \cite{Chivukula:2017qsi}. However, it is quite possible that the new mediators are at an energy scale that is beyond the reach of the LHC. A convenient method to study effects of high-scale physics in low-energy processes involves effective field theories (EFT) \cite{Petrov:2016azi}. If LFV happens to be at a scale $\Lambda$ that is beyond the reach of direct searches at the LHC, studies of LFV effects at the LHC can still be done using EFT methods. The low-energy effects of BSM physics generated at a UV scale $\Lambda$ can be characterized in terms of an effective Lagrangian $\mathcal{L}_\textrm{eff}$ containing terms of dimension $d \ge 5$ suppressed by appropriate powers of the NP scale $\Lambda$. In particular, at dimension 6 the following $SU(3)_C \times U(1)_\textrm{EM}$ invariant CLFV interactions are generated,
  \begin{align}
    \label{eq:4fermion_lagrangian}
    \mathcal{L}_\textrm{eff}^{(6)} \supset \frac{1}{\Lambda^2} \sum_{i,j,k,l,m,n} C_{ijkl}^{mn} \left ( \overline{\ell}_i \Gamma^m \ell_j \right ) \left ( \overline{q}_k \Gamma^n q_l \right ) + \textrm{h.c.} \, ,
  \end{align}
  where $i,j = 1,2,3$ label lepton generation, $k,l = 1,2,3$ label quark generation, $\Gamma^m$ denote the Dirac structure, and $C_{ijkl}^{mn}$ are Wilson coefficients. The operators in Eq.~\eqref{eq:4fermion_lagrangian} can be probed in a variety of ways, both at high~\cite{Black:2002wh,Han:2010sa,Arganda:2015ija,Cai:2015poa} and low energies~\cite{Hazard:2017udp,Hazard:2016fnc,Dreiner:2006gu,Daub:2012mu,Lindner:2016bgg,Davidson:2016edt,Crivellin:2013hpa,Crivellin:2017rmk,Celis:2014asa}.~\footnote{For an alternative approach to studying CLFV at fixed target experiments, see e.g.~\cite{Takeuchi:2017btl}.}

 The large parton luminosity for gluon-gluon interactions at high-energy $pp$ colliders, such as the LHC, implies that gluon-initiated processes might be prevalent there. However, the set of operators in Eq.~\eqref{eq:4fermion_lagrangian} does not contain gluon fields. The lowest order effective operator that is invariant under the SM gauge group $SU(3)_C \times SU(2)_L \times U(1)_Y$ that couples lepton and gluon fields appears at dimension eight,
  \begin{align}
    \mathcal{L}_\textrm{eff}^{(8)} = \frac{g_s^2}{\Lambda^4} \left [ Y_{ij} \overline{L}_L^{\, i} H \ell_R^{\, j} \, G \cdot G + \widetilde{Y}_{ij} \overline{L}_L^{\, i} H \ell_R^{\, j} \, G \cdot \widetilde{G} \right ] + \textrm{h.c.} \, ,
  \end{align}
  where $L_L^{\, i}$ represents the left-handed doublet lepton field with generation index $i$ in the gauge basis, $\ell_R^{\, i}$ is a right-handed lepton singlet field, and $H$ is a Higgs field. Gauge-invariant combinations of gluon fields are $G \cdot G \equiv G^a_{\mu \nu} G^{a \, \mu \nu}$, and $G \cdot \widetilde{G} \equiv G^a_{\mu \nu} \widetilde{G}^{a \, \mu \nu}$. Here $G^{a}_{\mu \nu}$ is a gluon field strength tensor and
  \begin{align}
    \widetilde{G}^a_{\mu \nu} = \frac{1}{2} \epsilon_{\mu \nu \alpha \beta}G^{a \, \alpha \beta}
  \end{align}
  is its dual. The couplings $Y_{ij} (\widetilde{Y}_{ij})$ are in general complex.  As spontaneous symmetry breaking leads to non-diagonal lepton mass matrices, their diagonalization will result in bi-unitary transformations of $Y_{ij} (\widetilde{Y}_{ij}) \to y_{ij} (\widetilde{y}_{ij})$.  Switching to a mass basis for lepton fields will then lead to LFV interactions of charged leptons $\ell^i$,
  \begin{align}
    \label{eq:gg_lagrangian}
    \mathcal{L}_\textrm{eff}^{(8)} = \frac{v g_s^2}{\sqrt{2} \Lambda^4} \left [ y_{ij} \overline{\ell}_L^{\, i} \ell_R^{\, j} \, G \cdot G + \widetilde{y}_{ij} \overline{\ell}_L^{\, i} \ell_R^{\, j} \, G \cdot \widetilde{G} \right ] + \textrm{h.c.} \, ,
  \end{align}
  where $v \sim 246$~GeV is the Higgs vacuum expectation value (VEV).

  For definiteness, we concentrate on the particular leptonic final state $\mu\tau$. In certain models of NP this final state might have the largest coupling to the new degrees of freedom, for instance due to the Cheng-Sher ansatz~\cite{Cheng:1987rs}. Additionally, final states with muons could be preferable from the point of view of experimental detection. For instance, searches for Higgs and $Z$-boson decays to $\mu \tau$ are common for studies of LFV at the LHC \cite{Arhrib:2012ax} by ATLAS \cite{Aad:2016blu} and CMS \cite{Khachatryan:2015kon,Sirunyan:2017xzt} collaborations.

\begin{figure}[t]
    \centering
    \begin{tabular}{m{0.025\textwidth} m{0.35\textwidth}}
      (a) & \includegraphics[width=0.325\textwidth]{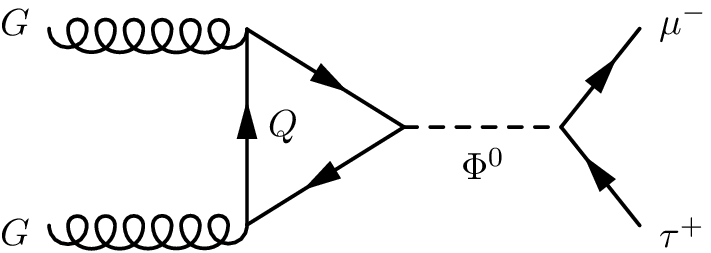} \\ \\
      (b) & \includegraphics[width=0.325\textwidth]{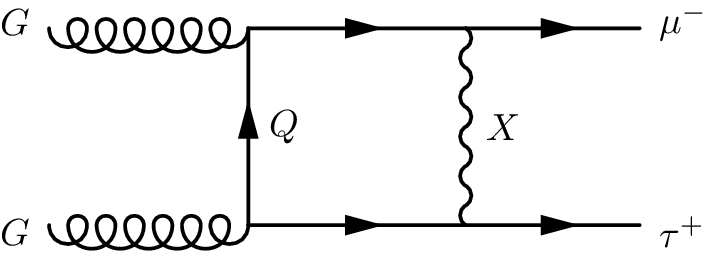}
    \end{tabular}
    \caption{Example Feynman diagrams which can generate the operators in Eq.~\eqref{eq:gg_operators}.}
    \label{fig:clfv_feynman}
  \end{figure}

  It is interesting to point out that the $v/\Lambda^4$ suppression of the operators in Eq.~\eqref{eq:gg_lagrangian} is not universal. Consider, for example, NP models where the effective coupling between gluons and leptons is generated after matching at one loop. This can be seen explicitly in two Higgs doublet models (2HDM) without natural flavor conservation with a heavy Higgs mediating CLFV as in Fig.~\ref{fig:clfv_feynman}(a) or in the case of CLFV mediated by a heavy scalar or vector lepto-quark with appropriate quantum numbers as in Fig.~\ref{fig:clfv_feynman}(b). Depending on the UV completion of the model, particles $Q$ and/or $\Phi^0/Z$ could belong to the NP or SM spectra. If for both particles, $m_Q \sim m_{\Phi^0} \sim \Lambda$ in Fig.~\ref{fig:clfv_feynman}(a) or $m_Q \sim m_{X} \sim \Lambda$ in Fig.~\ref{fig:clfv_feynman}(b), the overall scaling of the effective operators would be $\propto (16 \pi^2 \Lambda^4 / v)^{-1}$. Yet, if $Q$ is a standard model top quark, then at low energies one should expect the scaling of the effective operators to be $\propto (16 \pi^2 m_t \Lambda^2)^{-1}$. Such scaling of effective operators is standard in low-energy studies of lepton-flavor violation~\cite{Raidal:2008jk,Celis:2014asa,Petrov:2013vka}.  Finally, the large gluon luminosity of the LHC can affect the detection probabilities, selecting effective operators with explicit gluonic degrees of freedom, even though they could be suppressed by additional powers of $1/\Lambda$.

  It will therefore be appropriate, for the sake of a model-independent analysis, to introduce a set of dimension-full constants $\mathcal{C}^{\ell_1 \ell_2}_i$ that encode all effects of relevant Wilson coefficients and scales. Once these coefficients are constrained from the LHC data, we can then use the available constraints to discuss different ultraviolet completions (and thus interpretations) of the effective theory. The Lagrangian of Eq.~\eqref{eq:gg_lagrangian} then leads to the following interactions facilitating $\mu \tau$ production,
  \begin{align}
    \mathcal{L}_\textrm{eff} =  \sum_{i = 1}^4 \mathcal{C}_i^{\mu \tau} \mathcal{O}_i^{\mu \tau} + \textrm{h.c.} \, ,
  \end{align}
  where
  \begin{align}
    \label{eq:gg_operators}
    \renewcommand{\arraystretch}{1.25}
    \begin{array}{r l}
      \mathcal{O}_1^{\mu \tau} \hspace*{-0.25cm} &= \left ( \overline{\mu}_L \tau_R \right ) \, G \cdot G \, , \\
      \mathcal{O}_2^{\mu \tau} \hspace*{-0.25cm} &= \left ( \overline{\mu}_L \tau_R \right ) \, G \cdot \widetilde{G} \, , \\
      \mathcal{O}_3^{\mu \tau} \hspace*{-0.25cm} &= \left ( \overline{\mu}_R \tau_L \right ) \, G \cdot G \, , \\
      \mathcal{O}_4^{\mu \tau} \hspace*{-0.25cm} &= \left ( \overline{\mu}_R \tau_L \right ) \, G \cdot \widetilde{G} \, .
    \end{array}
  \end{align}

  The remainder of this letter proceeds as follows. In Sec.~\ref{sec:gluonic_operators}, we place constraints on the coefficients $\mathcal{C}_i^{\mu \tau}$ of gluonic operators. In Sec.~\ref{sec:ttbar_operators} we use those constraints to put limits on lepton-flavor violating couplings of top quarks that are difficult to constrain at low energy machines. We conclude in Sec.~\ref{sec:conclusion}.

  \section{LHC constraints on Gluonic Operators}
  \label{sec:gluonic_operators}

  \subsection{Event Selection}
  \label{sec:event_selection}

  At the 13 TeV LHC, the tau decays promptly into neutrinos and either an electron, a muon, or hadrons. The cleanest signal comes from leptonic $\tau$ decays, and since $\mu^+ \mu^-$ has a large SM background we will study the $\mu e$ final state. The leading backgrounds are then $W^+ W^-$ pair production, $Z^0 / \gamma^* \rightarrow \tau \tau$, and $t \overline{t}$ pair production. For this study we apply the basic cuts of Ref.~\cite{Han:2010sa}, which are reviewed below.

  For detector coverage and triggering, we require the transverse momentum $p_\textrm{T}$ and pseudorapidity $\eta$ to satisfy
  \begin{align}
    \label{eq:cut1}
    p_\textrm{T}^{\mu,e} > 20~\text{GeV} \;\;\; \text{ and } \;\;\; | \eta^{\mu,e} | < 2.5 \, ,
  \end{align}
  while vetoing events with a final state jet of $p_\textrm{T}^j > 50$~GeV and $| \eta^j | < 2.5$. For the signal, we anticipate the $\mu$ and $\tau$ to be back to back in the transverse plane with $p_\textrm{T}^\mu = p_\textrm{T}^\tau$ and with the decay products of the $\tau$ highly collimated. We therefore impose the additional requirements
  \begin{align}
    \label{eq:cut2}
    \renewcommand{\arraystretch}{1.25}
    \begin{array}{c}
      \delta \phi(p_\textrm{T}^\mu, p_\textrm{T}^e) > 2.5 \, , \;\;\;\;\; \delta \phi(p_\textrm{T}^\textrm{miss}, p_\textrm{T}^e) < 0.6 \, , \\
      \Delta p_T = p_\textrm{T}^\mu - p_\textrm{T}^e > 0 \, ,
    \end{array}
  \end{align}
  where $p_\textrm{T}^\textrm{miss}$ is the event's missing transverse momentum.

  The signal kinematics also allows us to approximately reconstruct the $\tau$. All of the missing energy in signal events is due to $\tau$ decay products, which gives
  \begin{align}
    \vec{p}_\textrm{T}^{\; \tau} = \vec{p}_\textrm{T}^{\; e} + \vec{p}_\textrm{T}^\textrm{ miss} \, .
  \end{align}
  From the expectation that the decay products of the $\tau$ will be highly collimated such that $p_z^e / p_z^\textrm{miss} \approx p_\textrm{T}^e / p_\textrm{T}^\textrm{miss}$. Thus, the longitudinal component of the $\tau$ momentum should be
  \begin{align}
    p_z^\tau \approx p_z^e \left ( 1 + \frac{p_\textrm{T}^\textrm{miss}}{p_\textrm{T}^e} \right ) \, .
  \end{align}
  Once the $\tau$'s 3-momentum is reconstructed the energy is $E_\tau^2 = \vec{p}_\tau^{\; 2} + m_\tau^2$. With the momentum of the $\tau$ fully reconstructed for signal events, we then require the invariant mass of the $\mu \tau$ system to satisfy
  \begin{align}
    \label{eq:cut3}
    M_{\mu \tau} > 250 \text{ GeV} \, ,
  \end{align}
  as the missing energy present in the backgrounds does not in general come from the decay of a single $\tau$.

  \subsection{Constraints}
  \label{sec:gg_constraints}

  To estimate constraints on the operators in Eq.~\eqref{eq:gg_operators} at the 13~TeV LHC, signal and background events were generated using {\sc MadGraph5}~\cite{Alwall:2014hca}. Showering and hadronization of these events, as well as decay of the $\tau$, was then performed using {\sc Pythia8}~\cite{Sjostrand:2006za, Sjostrand:2007gs}, while detector effects were simulated with {\sc Delphes}~\cite{deFavereau:2013fsa}. The signal model file was generated using {\sc FeynRules}~\cite{Alloul:2013bka}. Background and signal cross sections after applying successive cuts are shown in Table~\ref{tab:mutau_bgs}.
  \begin{table}[t]
    \begin{center}
      \renewcommand{\arraystretch}{1.15}
      \small
      \begin{tabular}{| c | c | c | c | c |}
        \hline \hline
        $\sigma$ (pb) & No cuts & $+$ Eq.~\eqref{eq:cut1} & $+$ Eq.~\eqref{eq:cut2} & $+$ Eq.~\eqref{eq:cut3} \\
        \hline
        $W W (\mu \tau)$ & 1.6 & 0.024 & 0.0044 & 0.0015 \\
        $W W (\mu e)$ & 1.6 & 0.35 & 0.014 & 0.0044 \\
        $Z/\gamma^* (\tau \tau)$ & 2400 & 1.7 & 0.26 & 0.00083 \\
        $t t (\mu \tau)$ & 12 & 0.043 & 0.0045 & 0.0019 \\
        $t t (\mu e)$ & 12 & 0.53 & 0.015 & 0.0081 \\
        \hline
        $\mathcal{O}_i^{\mu \tau}$ & 0.89 & 0.030 & 0.028 & 0.028 \\
        \hline \hline
      \end{tabular}
      \caption{Background and signal cross sections at the 13~TeV LHC. The signal cross section assumes the benchmark values of $C_i^{\mu \tau} = 4 \pi v \, g_s^2 / \sqrt{2} \Lambda^4$ with $\Lambda = 2$~TeV. Cross sections before cuts are given prior to $\tau$ decays.}
      \label{tab:mutau_bgs}
    \end{center}
  \end{table}
  Signal cross sections are calculated using the benchmark values of $\mathcal{C}_i^{\mu \tau} = 4 \pi v \, g_s^2 / \sqrt{2} \Lambda^4$ with $\Lambda = 2$~TeV. The running of the Wilson coefficients is assumed to be negligible. All operators are considered independently, and have the same cross section up to variations in their respective effective couplings.

  At 100~fb$^{-1}$ of integrated luminosity, we estimate the $2 \sigma$ confidence level (CL$_s$) exclusion limit and $5 \sigma$ log-likelihood (LL) discovery significance for $C_i^{\mu \tau}$ to be
  \begin{align}
    \left ( \mathcal{C}_i^{\mu \tau} \right )_{2 \sigma} &\approx \left ( 3300~\textrm{GeV} \right )^{-3} \, ,\\
    \left ( \mathcal{C}_i^{\mu \tau} \right )_{5 \sigma} &\approx \left ( 2900~\textrm{GeV} \right )^{-3} \, .
  \end{align}
  These estimates can be translated to general constraints on the NP scale $\Lambda$ of Eq.~\eqref{eq:gg_lagrangian} where $C_i^{\mu \tau} = 4 \pi v \, g_s^2 / \sqrt{2} \Lambda^4$. The values $y_{\mu \tau} = \widetilde{y}_{\mu \tau} = 4 \pi$ are chosen to push the perturbative limit of these operators in order to estimate the maximum sensitivity of the LHC to the various BSM scenarios discussed in Sec.~\ref{sec:introduction}. With these assumptions we find lower bounds on $\Lambda$ of
  \begin{align}
    \Lambda_{2 \sigma} &\approx 3000~\textrm{GeV} \, , \\
    \Lambda_{5 \sigma} &\approx 2800~\textrm{GeV} \, .
  \end{align}
  If we instead anticipate $y_{\mu \tau} = y_{\mu \tau} \sim \mathcal{O}(1)$, we find that the scale of these operators are constrained to be $\Lambda_{2 \sigma} \sim 1.6$~TeV. While we anticipate probing heavier NP scales as more data accumulates, models which generate the operators of Eq.~\eqref{eq:gg_operators} at a single UV scale have cross sections suppressed by $\Lambda^{-8}$ which limits the effectiveness of additional data on the ability to probe significantly higher scales at the LHC. A plot of the integrated luminosity at the 13~TeV LHC vs. $\Lambda$ is shown in Fig.~\ref{fig:luminosity_vs_cutoff} of Sec.~\ref{sec:ttbar_constraints}.
  
  The operators of Eq.~\eqref{eq:gg_operators} are in general also constrained by low energy experiments. For example, in Ref.~\cite{Petrov:2013vka} the authors present an analysis of constraints from limits on LFV tau decays to a muon and one or two hadrons. The results of their analysis, converted to the normalization used in this paper, are shown in Table~\ref{tab:tau_constraints}. The most stringent constraints, coming from $\tau \rightarrow \mu \pi^+ \pi^-$ for $\mathcal{O}_{1,3}^{\mu \tau}$ and $\tau \rightarrow \mu \eta$ for $\mathcal{O}_{2,4}^{\mu \tau}$, are $\Lambda_{1,3} \approx 1000$~GeV and $\Lambda_{2,4} \approx 830$~GeV.~\footnote{Alternative studies of similar processes offer differing estimates of the bounds from LFV tau decays (see e.g. Ref.~\cite{Celis:2014asa}), but these estimates generally fall well below the LHC's expected sensitivity.} These bounds are several times lower than the estimated sensitivity of the LHC with 100~fb$^{-1}$ of luminosity.
  
  \begin{table}[t]
    \centering
    \renewcommand{\arraystretch}{1.15}
    \begin{tabular}{| c | c | c |}
      \hline \hline
      Process & $C_{1,3}^{\mu \tau}$ (GeV$^{-3}$) & $\Lambda_{1,3}$ (GeV) \\
      \hline
      $\tau \rightarrow \mu \, \pi^+ \pi^-$ & $780^{-3}$ & 1000 \\
      $\tau \rightarrow \mu \, K^+ K^-$ & $700^{-3}$ & 950 \\
      \hline \hline
      Process & $C_{2,4}^{\mu \tau}$ (GeV$^{-3}$) & $\Lambda_{2,4}$ (GeV) \\
      \hline
      $\tau \rightarrow \mu \, \eta$ & $590^{-3}$ & 830 \\
      $\tau \rightarrow \mu \, \eta^\prime$ & $520^{-3}$ & 760 \\
      \hline
      \hline
    \end{tabular}
    \caption{Constraints on the coefficients of $\mathcal{O}_i^{\mu \tau}$ from $\tau$ decays, adapted from Ref.~\cite{Petrov:2013vka}. The constraints on $\Lambda$ are calculated using $C_i^{\mu \tau} = 4 \pi v \, g_s^2 / \sqrt{2} \Lambda^4_i$.}
    \label{tab:tau_constraints}
  \end{table}

  \section{LHC Constraints on $t \overline{t} \, \mu \tau$ Operators}
  \label{sec:ttbar_operators}

  Studies of $\mu \tau$ production at hadron colliders mediated by four-fermion operators have been performed~\cite{Han:2010sa}, with constraints on the NP scale obtained with the help of a single operator dominance hypothesis~\cite{Hazard:2016fnc}. They, however, did not examine operators that include top-quark fields. As we show below, these operators can be constrained by studying $gg\to\mu\tau$ processes.

  \subsection{Matching Conditions}
  \label{sec:matching_conditions}

  The $SU(3)_C \times U(1)_\textrm{EM}$ invariant Lagrangian contributing to $\mu \tau$ production contains
  \begin{align}
    \label{eq:4fermion_lagrangian2}
    \mathcal{L}_{\mu \tau}^{(6)} \supset \frac{1}{\Lambda^2} \sum_{i = 1}^4 C_i^{q \mu \tau} \mathcal{O}_i^{q \mu \tau} + \text{h.c.} \, ,
  \end{align}
  where
  \begin{align}
    \label{eq:4fermion_operators}
    \renewcommand{\arraystretch}{1.25}
    \begin{array}{r l}
      \mathcal{O}_1^{q \mu \tau} \hspace*{-0.25cm} &= \left ( \overline{\mu}_L \tau_R \right ) \left ( \overline{q}_L q_R \right ) \, , \\
      \mathcal{O}_2^{q \mu \tau} \hspace*{-0.25cm} &= \left ( \overline{\mu}_L \tau_R \right ) \left ( \overline{q}_R q_L \right ) \, , \\
      \mathcal{O}_3^{q \mu \tau} \hspace*{-0.25cm} &= \left ( \overline{\mu}_R \tau_L \right ) \left ( \overline{q}_L q_R \right ) \, , \\
      \mathcal{O}_4^{q \mu \tau} \hspace*{-0.25cm} &= \left ( \overline{\mu}_R \tau_L \right ) \left ( \overline{q}_R q_L \right ) \, .
    \end{array}
  \end{align}
  As noted in Section~\ref{sec:introduction}, these operators also generate the gluonic operators of Eq.~\eqref{eq:gg_operators} via SM quark loops in the diagrams represented in Fig.~\ref{fig:clfv_feynman}. Thus we can take advantage of the enhanced gluon luminosity at the LHC to probe these operators indirectly through gluon fusion production. A brief discussion of $SU(2)_L$ invariant operators generating those of Eq.~\eqref{eq:4fermion_operators} is given in \ref{sec:effective_operators}.

  The coefficients of the dimension 8 operators, $C_i^{\mu \tau}$, are related to the Wilson coefficients of the dimension 6 operators, $C_i^{q\mu\tau}$ by
  \begin{align}
    \label{eq:matching1}
    \mathcal{C}_{1,3}^{\mu \tau} &= \frac{g_s^2}{16 \pi^2} \frac{F_1(x)}{\Lambda^2 m_q} \left [ C_{1,3}^{q \mu \tau} + C_{2,4}^{q \mu \tau} \right ]_{C_{1,3}^{q \mu \tau} = C_{2,4}^{q \mu \tau}} \, , \\
    \label{eq:matching2}
    \mathcal{C}_{2,4}^{\mu \tau} &= \frac{i g_s^2}{16 \pi^2} \frac{F_2(x)}{\Lambda^2 m_q} \left [ C_{1,3}^{q \mu \tau} - C_{2,4}^{q \mu \tau} \right ]_{C_{1,3}^{q \mu \tau} = - C_{2,4}^{q \mu \tau}} \, ,
  \end{align}
  where $m_q$ is the mass of the quark running in the loop. Here $F(x)$ are functions of the parton center-of-momentum (CM) energy $\hat{s}$, and are given by
  \begin{align}
    \label{eq:form_factor1}
    F_1 (x) &= - \frac{x}{2} \left [ 4 + (4 x - 1) \ln^2 \left ( 1 - \frac{1}{2x} + \frac{\sqrt{1 - 4 x}}{2x} \right ) \right ] \, , \\
    \label{eq:form_factor2}
    F_2 (x) &= \frac{x}{2} \ln^2 \left ( 1 - \frac{1}{2x} + \frac{\sqrt{1 - 4 x}}{2x} \right ) \, ,
  \end{align}
  where $x \equiv m_q^2 / \hat{s}$. In the limit that $x \ll 1$, the functions $F(x)$ approach $m_q^2 / \hat{s}$, indicating that the contribution to $\mu\tau$ production from gluon fusion is dominated by the heaviest quark running in the loop. At LHC energies, provided only SM quarks contribute to this process, the top quark contribution is, therefore, expected to dominate.

  \subsection{Constraints}
  \label{sec:ttbar_constraints}

  Converting the results from Section~\ref{sec:gg_constraints}, we find, for 100~fb$^{-1}$ of integrated luminosity with the benchmark values $\left | C_i^{q \mu \tau} \right | = 4 \pi$, chosen again to be at the perturbative limit in order to estimate the maximum potential reach of the study, the $2 \sigma$ CL$_s$ exclusion limit and $5 \sigma$ LL discovery significance for the $G \cdot G$ operators $\mathcal{O}_{1,3}^{\mu \tau}$ to be
  \begin{align}
    \label{eq:dim6_constraint1}
    \Lambda_{2 \sigma} &\approx 3400~\textrm{GeV} \, , \\
    \label{eq:dim6_constraint2}
    \Lambda_{5 \sigma} &\approx 2900~\textrm{GeV} \, ,
  \end{align}
  while for the $G \cdot \widetilde{G}$ operators $\mathcal{O}_{2,4}^{\mu \tau}$ they are
  \begin{align}
    \label{eq:dim6_constraint3}
    \Lambda_{2 \sigma} &\approx 4100~\textrm{GeV} \, , \\
    \label{eq:dim6_constraint4}
    \Lambda_{5 \sigma} &\approx 3400~\textrm{GeV} \, .
  \end{align}
  Note that the energy scale $\Lambda$ here is the NP scale of the dimension 6 four-fermion operators of Eq.~\eqref{eq:4fermion_lagrangian2}.~\footnote{The gluon interactions with $\mu \tau$ are clearly non-local at LHC energies. To account for this, we average the full form factors (squared) of Eqs.~\eqref{eq:form_factor1} and \eqref{eq:form_factor2} by reconstructing the $\tau$ to obtain an approximate event-by-event $\hat{s}$.} For $|C_i^{q \mu \tau} | \sim \mathcal{O}(1)$, the constraints of Eqs.~(\ref{eq:dim6_constraint1}--\ref{eq:dim6_constraint4}) are estimated to be only $\Lambda_{2 \sigma} \sim 0.97$ (1.1)~TeV for the $G \cdot G$ ($G \cdot \widetilde{G}$) operators, which may be near the scale of validity for the EFT at the LHC. However, unlike the operators discussed in Sec.~\ref{sec:gluonic_operators}, the dimension 6 operator-induced cross-sections scale as $\Lambda^{-4}$ and therefore stand to benefit more from the accumulation of additional data.

  \begin{figure}[t]
    \centering
    \includegraphics[width=0.475\textwidth]{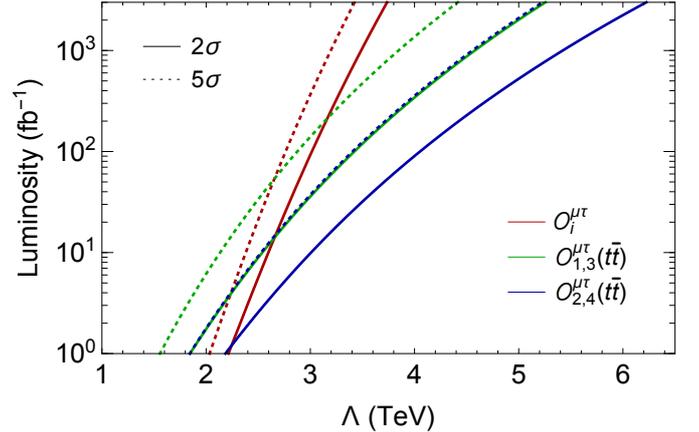}
    \caption{Luminosity goal at the 13 TeV LHC as a function of the NP scale $\Lambda$, choosing a value of $4 \pi$ for the dimensionless Wilson coefficients}. The solid (dashed) curves represent the $2\sigma$ CL$_s$ exclusion limit ($5\sigma$ LL discovery limit) on the luminosity required to rule out (discover) NP at the scale $\Lambda$.
    \label{fig:luminosity_vs_cutoff}
  \end{figure}
  
  Fig.~\ref{fig:luminosity_vs_cutoff} shows the luminosity required to set $2 \sigma$ CL$_s$ exclusion constraints and $5 \sigma$ LL discovery estimates as a function of the NP scale $\Lambda$. With 100~fb$^{-1}$ of integrated luminosity, the constraints on the operators in Eq.~\eqref{eq:gg_lagrangian} and those of Eq.~\eqref{eq:4fermion_operators} appear to be quite similar due to the explicit inclusion of loop suppression factors via the matching conditions of Eqs.~\eqref{eq:matching1} and \eqref{eq:matching2}. However, the LHC becomes increasingly sensitive to interactions via SM top quark loops as integrated luminosity is increased. As data at the LHC continues to accrue, experiments will become increasingly sensitive to NP that generates the operators of Eq.~\eqref{eq:4fermion_operators} and may be probed via gluonic processes. 
  
  The dimension 6 operators of Eq.~\eqref{eq:4fermion_operators} also induce at one loop couplings to the SM Higgs boson. LFV Higgs decays are stringently constrained by direct searches at the LHC, with the most recent result from the CMS collaboration at $\sqrt{s} = 13$~TeV constraining the branching ratio to be Br$(h \rightarrow \mu \tau) < 0.25\%$~\cite{Sirunyan:2017xzt}. In general, however, other operators may contribute to LFV Higgs couplings. Specifically, at dimension 6 the operators $(\overline{f}_L^i f_R^j) H (H^\dagger H)$ and $(\overline{f}_{L,R}^i \gamma^\mu f_{L,R}^j) (H^\dagger i \overleftrightarrow{D}_\mu H)$ induce direct LFV couplings to the Higgs boson, with potential interference between the various contributions. While a thorough study of the effects of these operators on Higgs decays is beyond the scope of this work, see e.g. Ref.~\cite{Harnik:2012pb} for an analysis of EFT-induced LFV couplings to the SM Higgs boson.

  \section{Conclusion}
  \label{sec:conclusion}

  In this letter we have examined CLFV processes initiated by gluon fusion at the $\sqrt{s} = 13$~TeV LHC. We have demonstrated that the gluon's enhanced parton luminosity can compensate for the increased suppression from dimension 8 operators relative to the less suppressed dimension 6 quark-induced CLFV processes. This allows one to indirectly probe models of NP mediating CLFV processes that may otherwise be inaccessible at LHC energies. The LHC has already collected nearly 100~fb$^{-1}$ of data at 13~TeV. We have estimated that this data can constrain the dimensionful coefficients of gluonic CLFV-inducing operators to be $\mathcal{C}_i^{\mu \tau} \gtrsim \left ( 3.3~\textrm{TeV} \right )^{-3}$.

  In addition, we have presented a study of such processes occurring through SM quark loops. In models where single operator dominance is expected, we have demonstrated that it is possible to constrain the CLFV coupling of leptons to top quarks through loop-induced gluon fusion. With 100~fb$^{-1}$ of data, we have estimated that the NP scale $\Lambda$ of $t \overline{t} \, \mu \tau$ couplings can be constrained to be $\Lambda \gtrsim 3.4 - 4.1$~TeV. This mechanism is especially important for models that predict an enhanced coupling to top quarks such as in certain 2HDMs with LFV. A future discovery of CLFV in the $\mu \tau$ final state at the LHC could be the first indication of preferential couplings to top quarks, and could be important in discriminating between the many models of CLFV.

  \section{Acknowledgements}
  \label{sec:acknowledgements}

  This work has been supported in part by the U.S. Department of Energy under contract DE-SC0007983 and by the National Science Foundation under grant PHY-1460853 under the auspices of WSU Research Experience for Undergraduates program.

  \begin{appendix}

    \section{$SU(2)_L$ Invariant Operators}
    \label{sec:effective_operators}

    The $SU(3)_C \times U(1)_\textrm{EM}$ invariant operators of Eq.~\eqref{eq:4fermion_operators} arise from $SU(2)_L$ invariant forms. Specifically, $\mathcal{O}_1^{q \mu \tau}$ is contained in the dimension 6 operator
    \begin{align}
      \label{eq:su2_dim6}
      \mathcal{O}^{LeQu} = ( \overline{L}_L^{\, i} e_R^{\, j} ) \, \epsilon \, ( \overline{Q}_L^{\, k} u_R^{\, l} ) \, ,
    \end{align}
    where the antisymmetric tensor $\epsilon$ contracts the suppressed $SU(2)_L$ indices, and $\mathcal{O}_4^{q \mu \tau}$ is included in its Hermitian conjugate. Conversely, operators $\mathcal{O}_{2,3}^{q \mu \tau}$ are first generated at dimension 8,
    \begin{align}
      \label{eq:su2_dim8}
      \mathcal{O}^{LHeuHQ} = ( [ \overline{L}_L^{\, i} H ] e_R^{\, j} ) ( \overline{u}_R^{\, k} [ H^T i \sigma_2 Q_L^{\, l}] )
    \end{align}
    and its Hermitian conjugate. Operators $\mathcal{O}_{1,4}^{q \mu \tau}$ can also be generated at dimension 8 without the associated charged current interactions of Eq.~\eqref{eq:su2_dim6}.

    Because only two of the operators listed in Eq.~\eqref{eq:4fermion_operators} appear at dimension 6 in an $SU(2)_L$ invariant form, there is in general no reason to expect the coefficients of these operators to be similar in value, as required by the matching conditions given in Eqs.~\eqref{eq:matching1} and \eqref{eq:matching2}. We should then generally expect a mixing of the $G \cdot G$ and $G \cdot \widetilde{G}$ production mechanisms. However, Ref.~\cite{Potter:2012yv} has demonstrated that the contributions from these operators can not be distinguished by a study of lepton pair production alone, and thus one should consider the limits presented in Section~\ref{sec:ttbar_constraints} as estimates of the upper bounds on such processes. We postpone a more complete discussion of the four-lepton operators for a future, more detailed, analysis, where one can expect the scale of $\mathcal{O}_{2,3}^{t \mu \tau}$ to be more weakly constrained than $\mathcal{O}_{1,4}^{t \mu \tau}$.

  \end{appendix}

  \section*{References}
  \label{sec:references}

  \bibliographystyle{elsarticle-num}
  \bibliography{mybibfile}

\begin{thebibliography}{10}
\expandafter\ifx\csname url\endcsname\relax
  \def\url#1{\texttt{#1}}\fi
\expandafter\ifx\csname urlprefix\endcsname\relax\def\urlprefix{URL }\fi
\expandafter\ifx\csname href\endcsname\relax
  \def\href#1#2{#2} \def\path#1{#1}\fi

\bibitem{Raidal:2008jk}
M.~Raidal, et~al., {Flavour physics of leptons and dipole moments}, Eur. Phys.
  J. C57 (2008) 13--182.
\newblock \href {http://arxiv.org/abs/0801.1826} {\path{arXiv:0801.1826}},
  \href {http://dx.doi.org/10.1140/epjc/s10052-008-0715-2}
  {\path{doi:10.1140/epjc/s10052-008-0715-2}}.

\bibitem{Aaij:2014ora}
R.~Aaij, et~al., {Test of lepton universality using $B^{+}\rightarrow
  K^{+}\ell^{+}\ell^{-}$ decays}, Phys. Rev. Lett. 113 (2014) 151601.
\newblock \href {http://arxiv.org/abs/1406.6482} {\path{arXiv:1406.6482}},
  \href {http://dx.doi.org/10.1103/PhysRevLett.113.151601}
  {\path{doi:10.1103/PhysRevLett.113.151601}}.

\bibitem{Aaij:2015yra}
R.~Aaij, et~al., {Measurement of the ratio of branching fractions
  $\mathcal{B}(\bar{B}^0 \to
  D^{*+}\tau^{-}\bar{\nu}_{\tau})/\mathcal{B}(\bar{B}^0 \to
  D^{*+}\mu^{-}\bar{\nu}_{\mu})$}, Phys. Rev. Lett. 115~(11) (2015) 111803,
  [Erratum: Phys. Rev. Lett.115,no.15,159901(2015)].
\newblock \href {http://arxiv.org/abs/1506.08614} {\path{arXiv:1506.08614}},
  \href {http://dx.doi.org/10.1103/PhysRevLett.115.159901,
  10.1103/PhysRevLett.115.111803} {\path{doi:10.1103/PhysRevLett.115.159901,
  10.1103/PhysRevLett.115.111803}}.

\bibitem{Aaij:2017vbb}
R.~Aaij, et~al., {Test of lepton universality with $B^{0} \rightarrow
  K^{*0}\ell^{+}\ell^{-}$ decays}, JHEP 08 (2017) 055.
\newblock \href {http://arxiv.org/abs/1705.05802} {\path{arXiv:1705.05802}},
  \href {http://dx.doi.org/10.1007/JHEP08(2017)055}
  {\path{doi:10.1007/JHEP08(2017)055}}.

\bibitem{Lees:2013uzd}
J.~P. Lees, et~al., {Measurement of an Excess of $\bar{B} \to D^{(*)}\tau^-
  \bar{\nu}_\tau$ Decays and Implications for Charged Higgs Bosons}, Phys. Rev.
  D88~(7) (2013) 072012.
\newblock \href {http://arxiv.org/abs/1303.0571} {\path{arXiv:1303.0571}},
  \href {http://dx.doi.org/10.1103/PhysRevD.88.072012}
  {\path{doi:10.1103/PhysRevD.88.072012}}.

\bibitem{Huschle:2015rga}
M.~Huschle, et~al., {Measurement of the branching ratio of $\bar{B} \to
  D^{(\ast)} \tau^- \bar{\nu}_\tau$ relative to $\bar{B} \to D^{(\ast)} \ell^-
  \bar{\nu}_\ell$ decays with hadronic tagging at Belle}, Phys. Rev. D92~(7)
  (2015) 072014.
\newblock \href {http://arxiv.org/abs/1507.03233} {\path{arXiv:1507.03233}},
  \href {http://dx.doi.org/10.1103/PhysRevD.92.072014}
  {\path{doi:10.1103/PhysRevD.92.072014}}.

\bibitem{Wehle:2016yoi}
S.~Wehle, et~al., {Lepton-Flavor-Dependent Angular Analysis of $B\to K^\ast
  \ell^+\ell^-$}, Phys. Rev. Lett. 118~(11) (2017) 111801.
\newblock \href {http://arxiv.org/abs/1612.05014} {\path{arXiv:1612.05014}},
  \href {http://dx.doi.org/10.1103/PhysRevLett.118.111801}
  {\path{doi:10.1103/PhysRevLett.118.111801}}.

\bibitem{Descotes-Genon:2013wba}
S.~Descotes-Genon, J.~Matias, J.~Virto, {Understanding the $B\to K^*\mu^+\mu^-$
  Anomaly}, Phys. Rev. D88 (2013) 074002.
\newblock \href {http://arxiv.org/abs/1307.5683} {\path{arXiv:1307.5683}},
  \href {http://dx.doi.org/10.1103/PhysRevD.88.074002}
  {\path{doi:10.1103/PhysRevD.88.074002}}.

\bibitem{Altmannshofer:2013foa}
W.~Altmannshofer, D.~M. Straub, {New Physics in $B \to K^*\mu\mu$?}, Eur. Phys.
  J. C73 (2013) 2646.
\newblock \href {http://arxiv.org/abs/1308.1501} {\path{arXiv:1308.1501}},
  \href {http://dx.doi.org/10.1140/epjc/s10052-013-2646-9}
  {\path{doi:10.1140/epjc/s10052-013-2646-9}}.

\bibitem{Gauld:2013qja}
R.~Gauld, F.~Goertz, U.~Haisch, {An explicit Z'-boson explanation of the $B \to
  K^* \mu^+ \mu^-$ anomaly}, JHEP 01 (2014) 069.
\newblock \href {http://arxiv.org/abs/1310.1082} {\path{arXiv:1310.1082}},
  \href {http://dx.doi.org/10.1007/JHEP01(2014)069}
  {\path{doi:10.1007/JHEP01(2014)069}}.

\bibitem{Datta:2013kja}
A.~Datta, M.~Duraisamy, D.~Ghosh, {Explaining the $B \to K^\ast \mu^+ \mu^-$
  data with scalar interactions}, Phys. Rev. D89~(7) (2014) 071501.
\newblock \href {http://arxiv.org/abs/1310.1937} {\path{arXiv:1310.1937}},
  \href {http://dx.doi.org/10.1103/PhysRevD.89.071501}
  {\path{doi:10.1103/PhysRevD.89.071501}}.

\bibitem{Buras:2013dea}
A.~J. Buras, F.~De~Fazio, J.~Girrbach, {331 models facing new $b \to s\mu^+
  \mu^-$ data}, JHEP 02 (2014) 112.
\newblock \href {http://arxiv.org/abs/1311.6729} {\path{arXiv:1311.6729}},
  \href {http://dx.doi.org/10.1007/JHEP02(2014)112}
  {\path{doi:10.1007/JHEP02(2014)112}}.

\bibitem{Altmannshofer:2014cfa}
W.~Altmannshofer, S.~Gori, M.~Pospelov, I.~Yavin, {Quark flavor transitions in
  $L_\mu-L_\tau$ models}, Phys. Rev. D89 (2014) 095033.
\newblock \href {http://arxiv.org/abs/1403.1269} {\path{arXiv:1403.1269}},
  \href {http://dx.doi.org/10.1103/PhysRevD.89.095033}
  {\path{doi:10.1103/PhysRevD.89.095033}}.

\bibitem{Sakaki:2013bfa}
Y.~Sakaki, M.~Tanaka, A.~Tayduganov, R.~Watanabe, {Testing leptoquark models in
  $\bar B \to D^{(*)} \tau \bar\nu$}, Phys. Rev. D88~(9) (2013) 094012.
\newblock \href {http://arxiv.org/abs/1309.0301} {\path{arXiv:1309.0301}},
  \href {http://dx.doi.org/10.1103/PhysRevD.88.094012}
  {\path{doi:10.1103/PhysRevD.88.094012}}.

\bibitem{Buttazzo:2017ixm}
D.~Buttazzo, A.~Greljo, G.~Isidori, D.~Marzocca, {B-physics anomalies: a guide
  to combined explanations}, JHEP 11 (2017) 044.
\newblock \href {http://arxiv.org/abs/1706.07808} {\path{arXiv:1706.07808}},
  \href {http://dx.doi.org/10.1007/JHEP11(2017)044}
  {\path{doi:10.1007/JHEP11(2017)044}}.

\bibitem{Celis:2015ara}
A.~Celis, J.~Fuentes-Martin, M.~Jung, H.~Serodio, {Family nonuniversal Z'
  models with protected flavor-changing interactions}, Phys. Rev. D92~(1)
  (2015) 015007.
\newblock \href {http://arxiv.org/abs/1505.03079} {\path{arXiv:1505.03079}},
  \href {http://dx.doi.org/10.1103/PhysRevD.92.015007}
  {\path{doi:10.1103/PhysRevD.92.015007}}.

\bibitem{Alonso:2015sja}
R.~Alonso, B.~Grinstein, J.~Martin~Camalich, {Lepton universality violation and
  lepton flavor conservation in $B$-meson decays}, JHEP 10 (2015) 184.
\newblock \href {http://arxiv.org/abs/1505.05164} {\path{arXiv:1505.05164}},
  \href {http://dx.doi.org/10.1007/JHEP10(2015)184}
  {\path{doi:10.1007/JHEP10(2015)184}}.

\bibitem{Glashow:2014iga}
S.~L. Glashow, D.~Guadagnoli, K.~Lane, {Lepton Flavor Violation in $B$
  Decays?}, Phys. Rev. Lett. 114 (2015) 091801.
\newblock \href {http://arxiv.org/abs/1411.0565} {\path{arXiv:1411.0565}},
  \href {http://dx.doi.org/10.1103/PhysRevLett.114.091801}
  {\path{doi:10.1103/PhysRevLett.114.091801}}.

\bibitem{Bhattacharya:2014wla}
B.~Bhattacharya, A.~Datta, D.~London, S.~Shivashankara, {Simultaneous
  Explanation of the $R_K$ and $R(D^{(*)})$ Puzzles}, Phys. Lett. B742 (2015)
  370--374.
\newblock \href {http://arxiv.org/abs/1412.7164} {\path{arXiv:1412.7164}},
  \href {http://dx.doi.org/10.1016/j.physletb.2015.02.011}
  {\path{doi:10.1016/j.physletb.2015.02.011}}.

\bibitem{Bhattacharya:2016mcc}
B.~Bhattacharya, A.~Datta, J.-P. Guvin, D.~London, R.~Watanabe, {Simultaneous
  Explanation of the $R_K$ and $R_{D^{(*)}}$ Puzzles: a Model Analysis}, JHEP
  01 (2017) 015.
\newblock \href {http://arxiv.org/abs/1609.09078} {\path{arXiv:1609.09078}},
  \href {http://dx.doi.org/10.1007/JHEP01(2017)015}
  {\path{doi:10.1007/JHEP01(2017)015}}.

\bibitem{Alok:2017jgr}
A.~K. Alok, B.~Bhattacharya, D.~Kumar, J.~Kumar, D.~London, S.~U. Sankar, {New
  physics in $b \rightarrow s \mu^+ \mu^-$: Distinguishing models through
  CP-violating effects}, Phys. Rev. D96~(1) (2017) 015034.
\newblock \href {http://arxiv.org/abs/1703.09247} {\path{arXiv:1703.09247}},
  \href {http://dx.doi.org/10.1103/PhysRevD.96.015034}
  {\path{doi:10.1103/PhysRevD.96.015034}}.

\bibitem{Alok:2017sui}
A.~K. Alok, B.~Bhattacharya, A.~Datta, D.~Kumar, J.~Kumar, D.~London, {New
  Physics in $b \to s \mu^+ \mu^-$ after the Measurement of $R_{K^*}$}, Phys.
  Rev. D96~(9) (2017) 095009.
\newblock \href {http://arxiv.org/abs/1704.07397} {\path{arXiv:1704.07397}},
  \href {http://dx.doi.org/10.1103/PhysRevD.96.095009}
  {\path{doi:10.1103/PhysRevD.96.095009}}.

\bibitem{Altmannshofer:2017yso}
W.~Altmannshofer, P.~Stangl, D.~M. Straub, {Interpreting Hints for Lepton
  Flavor Universality Violation}, Phys. Rev. D96~(5) (2017) 055008.
\newblock \href {http://arxiv.org/abs/1704.05435} {\path{arXiv:1704.05435}},
  \href {http://dx.doi.org/10.1103/PhysRevD.96.055008}
  {\path{doi:10.1103/PhysRevD.96.055008}}.

\bibitem{Crivellin:2017zlb}
A.~Crivellin, D.~Mueller, T.~Ota, {Simultaneous explanation of R(D$^{(*)}$) and
  $b \to s \mu^+\mu^-$: the last scalar leptoquarks standing}, JHEP 09 (2017)
  040.
\newblock \href {http://arxiv.org/abs/1703.09226} {\path{arXiv:1703.09226}},
  \href {http://dx.doi.org/10.1007/JHEP09(2017)040}
  {\path{doi:10.1007/JHEP09(2017)040}}.

\bibitem{Iguro:2017ysu}
S.~Iguro, K.~Tobe, {$R(D^{(*)})$ in a general two Higgs doublet model}, Nucl.
  Phys. B925 (2017) 560--606.
\newblock \href {http://arxiv.org/abs/1708.06176} {\path{arXiv:1708.06176}},
  \href {http://dx.doi.org/10.1016/j.nuclphysb.2017.10.014}
  {\path{doi:10.1016/j.nuclphysb.2017.10.014}}.

\bibitem{Iguro:2018qzf}
S.~Iguro, Y.~Omura, {Status of the semileptonic $B$ decays and muon g-2 in
  general 2HDMs with right-handed neutrinos}\href
  {http://arxiv.org/abs/1802.01732} {\path{arXiv:1802.01732}}.

\bibitem{Chivukula:2017qsi}
R.~S. Chivukula, J.~Isaacson, K.~A. Mohan, D.~Sengupta, E.~H. Simmons, {$R_K$
  anomalies and simplified limits on $Z'$ models at the LHC}, Phys. Rev.
  D96~(7) (2017) 075012.
\newblock \href {http://arxiv.org/abs/1706.06575} {\path{arXiv:1706.06575}},
  \href {http://dx.doi.org/10.1103/PhysRevD.96.075012}
  {\path{doi:10.1103/PhysRevD.96.075012}}.

\bibitem{Petrov:2016azi}
A.~A. Petrov, A.~E. Blechman,
  \href{http://www.worldscientific.com/worldscibooks/10.1142/8619}{{Effective
  Field Theories}}, World Scientific Publishing, 2016.
\newline\urlprefix\url{http://www.worldscientific.com/worldscibooks/10.1142/8619}

\bibitem{Black:2002wh}
D.~Black, T.~Han, H.-J. He, M.~Sher, {$\tau - \mu$ flavor violation as a probe
  of the scale of new physics}, Phys. Rev. D66 (2002) 053002.
\newblock \href {http://arxiv.org/abs/hep-ph/0206056}
  {\path{arXiv:hep-ph/0206056}}, \href
  {http://dx.doi.org/10.1103/PhysRevD.66.053002}
  {\path{doi:10.1103/PhysRevD.66.053002}}.

\bibitem{Han:2010sa}
T.~Han, I.~Lewis, M.~Sher, {Mu-Tau Production at Hadron Colliders}, JHEP 03
  (2010) 090.
\newblock \href {http://arxiv.org/abs/1001.0022} {\path{arXiv:1001.0022}},
  \href {http://dx.doi.org/10.1007/JHEP03(2010)090}
  {\path{doi:10.1007/JHEP03(2010)090}}.

\bibitem{Arganda:2015ija}
E.~Arganda, M.~J. Herrero, X.~Marcano, C.~Weiland, {Exotic $\mu\tau jj$ events
  from heavy ISS neutrinos at the LHC}, Phys. Lett. B752 (2016) 46--50.
\newblock \href {http://arxiv.org/abs/1508.05074} {\path{arXiv:1508.05074}},
  \href {http://dx.doi.org/10.1016/j.physletb.2015.11.013}
  {\path{doi:10.1016/j.physletb.2015.11.013}}.

\bibitem{Cai:2015poa}
Y.~Cai, M.~A. Schmidt, {A Case Study of the Sensitivity to LFV Operators with
  Precision Measurements and the LHC}, JHEP 02 (2016) 176.
\newblock \href {http://arxiv.org/abs/1510.02486} {\path{arXiv:1510.02486}},
  \href {http://dx.doi.org/10.1007/JHEP02(2016)176}
  {\path{doi:10.1007/JHEP02(2016)176}}.

\bibitem{Hazard:2017udp}
D.~E. Hazard, A.~A. Petrov, {Radiative lepton flavor violating B, D, and K
  decays}\href {http://arxiv.org/abs/1711.05314} {\path{arXiv:1711.05314}}.

\bibitem{Hazard:2016fnc}
D.~E. Hazard, A.~A. Petrov, {Lepton flavor violating quarkonium decays}, Phys.
  Rev. D94~(7) (2016) 074023.
\newblock \href {http://arxiv.org/abs/1607.00815} {\path{arXiv:1607.00815}},
  \href {http://dx.doi.org/10.1103/PhysRevD.94.074023}
  {\path{doi:10.1103/PhysRevD.94.074023}}.

\bibitem{Dreiner:2006gu}
H.~K. Dreiner, M.~Kramer, B.~O'Leary, {Bounds on R-parity violating
  supersymmetric couplings from leptonic and semi-leptonic meson decays}, Phys.
  Rev. D75 (2007) 114016.
\newblock \href {http://arxiv.org/abs/hep-ph/0612278}
  {\path{arXiv:hep-ph/0612278}}, \href
  {http://dx.doi.org/10.1103/PhysRevD.75.114016}
  {\path{doi:10.1103/PhysRevD.75.114016}}.

\bibitem{Daub:2012mu}
J.~T. Daub, H.~K. Dreiner, C.~Hanhart, B.~Kubis, U.~G. Meissner, {Improving the
  Hadron Physics of Non-Standard-Model Decays: Example Bounds on R-parity
  Violation}, JHEP 01 (2013) 179.
\newblock \href {http://arxiv.org/abs/1212.4408} {\path{arXiv:1212.4408}},
  \href {http://dx.doi.org/10.1007/JHEP01(2013)179}
  {\path{doi:10.1007/JHEP01(2013)179}}.

\bibitem{Lindner:2016bgg}
M.~Lindner, M.~Platscher, F.~S. Queiroz, {A Call for New Physics : The Muon
  Anomalous Magnetic Moment and Lepton Flavor Violation}, Phys. Rep.\href
  {http://arxiv.org/abs/1610.06587} {\path{arXiv:1610.06587}}, \href
  {http://dx.doi.org/10.1016/j.physrep.2017.12.001}
  {\path{doi:10.1016/j.physrep.2017.12.001}}.

\bibitem{Davidson:2016edt}
S.~Davidson, {$\mu \rightarrow e \gamma $ and matching at $m_W$}, Eur. Phys. J.
  C76~(7) (2016) 370.
\newblock \href {http://arxiv.org/abs/1601.07166} {\path{arXiv:1601.07166}},
  \href {http://dx.doi.org/10.1140/epjc/s10052-016-4207-5}
  {\path{doi:10.1140/epjc/s10052-016-4207-5}}.

\bibitem{Crivellin:2013hpa}
A.~Crivellin, S.~Najjari, J.~Rosiek, {Lepton Flavor Violation in the Standard
  Model with general Dimension-Six Operators}, JHEP 04 (2014) 167.
\newblock \href {http://arxiv.org/abs/1312.0634} {\path{arXiv:1312.0634}},
  \href {http://dx.doi.org/10.1007/JHEP04(2014)167}
  {\path{doi:10.1007/JHEP04(2014)167}}.

\bibitem{Crivellin:2017rmk}
A.~Crivellin, S.~Davidson, G.~M. Pruna, A.~Signer, {Renormalisation-group
  improved analysis of $\mu\to e$ processes in a systematic
  effective-field-theory approach}, JHEP 05 (2017) 117.
\newblock \href {http://arxiv.org/abs/1702.03020} {\path{arXiv:1702.03020}},
  \href {http://dx.doi.org/10.1007/JHEP05(2017)117}
  {\path{doi:10.1007/JHEP05(2017)117}}.

\bibitem{Celis:2014asa}
A.~Celis, V.~Cirigliano, E.~Passemar, {Model-discriminating power of lepton
  flavor violating $\tau$ decays}, Phys. Rev. D89~(9) (2014) 095014.
\newblock \href {http://arxiv.org/abs/1403.5781} {\path{arXiv:1403.5781}},
  \href {http://dx.doi.org/10.1103/PhysRevD.89.095014}
  {\path{doi:10.1103/PhysRevD.89.095014}}.

\bibitem{Takeuchi:2017btl}
M.~Takeuchi, Y.~Uesaka, M.~Yamanaka, {Higgs mediated CLFV processes $\mu N (eN
  ) \to \tau X$ via gluon operators}, Phys. Lett. B772 (2017) 279--282.
\newblock \href {http://arxiv.org/abs/1705.01059} {\path{arXiv:1705.01059}},
  \href {http://dx.doi.org/10.1016/j.physletb.2017.06.054}
  {\path{doi:10.1016/j.physletb.2017.06.054}}.

\bibitem{Cheng:1987rs}
T.~P. Cheng, M.~Sher, {Mass Matrix Ansatz and Flavor Nonconservation in Models
  with Multiple Higgs Doublets}, Phys. Rev. D35 (1987) 3484.
\newblock \href {http://dx.doi.org/10.1103/PhysRevD.35.3484}
  {\path{doi:10.1103/PhysRevD.35.3484}}.

\bibitem{Arhrib:2012ax}
A.~Arhrib, Y.~Cheng, O.~C.~W. Kong, {Comprehensive analysis on lepton flavor
  violating Higgs boson to $\mu^\mp \tau^\pm$ decay in supersymmetry without
  $R$ parity}, Phys. Rev. D87~(1) (2013) 015025.
\newblock \href {http://arxiv.org/abs/1210.8241} {\path{arXiv:1210.8241}},
  \href {http://dx.doi.org/10.1103/PhysRevD.87.015025}
  {\path{doi:10.1103/PhysRevD.87.015025}}.

\bibitem{Aad:2016blu}
G.~Aad, et~al., {Search for lepton-flavour-violating decays of the Higgs and
  $Z$ bosons with the ATLAS detector}, Eur. Phys. J. C77~(2) (2017) 70.
\newblock \href {http://arxiv.org/abs/1604.07730} {\path{arXiv:1604.07730}},
  \href {http://dx.doi.org/10.1140/epjc/s10052-017-4624-0}
  {\path{doi:10.1140/epjc/s10052-017-4624-0}}.

\bibitem{Khachatryan:2015kon}
V.~Khachatryan, et~al., {Search for Lepton-Flavour-Violating Decays of the
  Higgs Boson}, Phys. Lett. B749 (2015) 337--362.
\newblock \href {http://arxiv.org/abs/1502.07400} {\path{arXiv:1502.07400}},
  \href {http://dx.doi.org/10.1016/j.physletb.2015.07.053}
  {\path{doi:10.1016/j.physletb.2015.07.053}}.

\bibitem{Sirunyan:2017xzt}
A.~M. Sirunyan, et~al., {Search for lepton flavour violating decays of the
  Higgs boson to $\mu\tau$ and e$\tau$ in proton-proton collisions at
  $\sqrt{s}=$ 13 TeV}\href {http://arxiv.org/abs/1712.07173}
  {\path{arXiv:1712.07173}}.

\bibitem{Petrov:2013vka}
A.~A. Petrov, D.~V. Zhuridov, {Lepton flavor-violating transitions in effective
  field theory and gluonic operators}, Phys. Rev. D89~(3) (2014) 033005.
\newblock \href {http://arxiv.org/abs/1308.6561} {\path{arXiv:1308.6561}},
  \href {http://dx.doi.org/10.1103/PhysRevD.89.033005}
  {\path{doi:10.1103/PhysRevD.89.033005}}.

\bibitem{Alwall:2014hca}
J.~Alwall, R.~Frederix, S.~Frixione, V.~Hirschi, F.~Maltoni, O.~Mattelaer,
  H.~S. Shao, T.~Stelzer, P.~Torrielli, M.~Zaro, {The automated computation of
  tree-level and next-to-leading order differential cross sections, and their
  matching to parton shower simulations}, JHEP 07 (2014) 079.
\newblock \href {http://arxiv.org/abs/1405.0301} {\path{arXiv:1405.0301}},
  \href {http://dx.doi.org/10.1007/JHEP07(2014)079}
  {\path{doi:10.1007/JHEP07(2014)079}}.

\bibitem{Sjostrand:2006za}
T.~Sjostrand, S.~Mrenna, P.~Z. Skands, {PYTHIA 6.4 Physics and Manual}, JHEP 05
  (2006) 026.
\newblock \href {http://arxiv.org/abs/hep-ph/0603175}
  {\path{arXiv:hep-ph/0603175}}, \href
  {http://dx.doi.org/10.1088/1126-6708/2006/05/026}
  {\path{doi:10.1088/1126-6708/2006/05/026}}.

\bibitem{Sjostrand:2007gs}
T.~Sjostrand, S.~Mrenna, P.~Z. Skands, {A Brief Introduction to PYTHIA 8.1},
  Comput. Phys. Commun. 178 (2008) 852--867.
\newblock \href {http://arxiv.org/abs/0710.3820} {\path{arXiv:0710.3820}},
  \href {http://dx.doi.org/10.1016/j.cpc.2008.01.036}
  {\path{doi:10.1016/j.cpc.2008.01.036}}.

\bibitem{deFavereau:2013fsa}
J.~de~Favereau, C.~Delaere, P.~Demin, A.~Giammanco, V.~Lema�tre, A.~Mertens,
  M.~Selvaggi, {DELPHES 3, A modular framework for fast simulation of a generic
  collider experiment}, JHEP 02 (2014) 057.
\newblock \href {http://arxiv.org/abs/1307.6346} {\path{arXiv:1307.6346}},
  \href {http://dx.doi.org/10.1007/JHEP02(2014)057}
  {\path{doi:10.1007/JHEP02(2014)057}}.

\bibitem{Alloul:2013bka}
A.~Alloul, N.~D. Christensen, C.~Degrande, C.~Duhr, B.~Fuks, {FeynRules 2.0 - A
  complete toolbox for tree-level phenomenology}, Comput. Phys. Commun. 185
  (2014) 2250--2300.
\newblock \href {http://arxiv.org/abs/1310.1921} {\path{arXiv:1310.1921}},
  \href {http://dx.doi.org/10.1016/j.cpc.2014.04.012}
  {\path{doi:10.1016/j.cpc.2014.04.012}}.

\bibitem{Harnik:2012pb}
R.~Harnik, J.~Kopp, J.~Zupan, {Flavor Violating Higgs Decays}, JHEP 03 (2013)
  026.
\newblock \href {http://arxiv.org/abs/1209.1397} {\path{arXiv:1209.1397}},
  \href {http://dx.doi.org/10.1007/JHEP03(2013)026}
  {\path{doi:10.1007/JHEP03(2013)026}}.

\bibitem{Potter:2012yv}
H.~Potter, G.~Valencia, {Probing lepton gluonic couplings at the LHC}, Phys.
  Lett. B713 (2012) 95--98.
\newblock \href {http://arxiv.org/abs/1202.1780} {\path{arXiv:1202.1780}},
  \href {http://dx.doi.org/10.1016/j.physletb.2012.05.052}
  {\path{doi:10.1016/j.physletb.2012.05.052}}.

\end{thebibliography}

\end{document}